\documentclass{PoS}

\title{Sharing lattices throughout the world: an ILDG status report}

\ShortTitle{ILDG Status Report}

\author{\speaker{Carleton DeTar}\\
        Physics Department, University of Utah, Salt Lake City, UT 84112, USA\\
        E-mail: \email{detar@physics.utah.edu}}

\abstract{The international lattice data grid, a system for sharing
  gauge configuration files throughout the world, is operational for
  the most part.  I give a status report, give some pointers on
  accessing lattice files, and highlight some of the available gauge
  configurations.}

\FullConference{The XXV International Symposium on Lattice Field Theory\\
		 July 30 - August 4 2007\\
		 Regensburg, Germany}

\begin{document}

\section{Introduction}

The rapid expansion of internet services in the 1990's made it easy to
publish gauge configuration files for general public access.  The US
Gauge Connection at NERSC was established in 1997 for this
purpose\cite{NERSC}.  It was simple and effective, but it was not the
ideal solution for broad international use.  The development of a
truly international lattice data grid (ILDG) began some years later
(2002) as a result of discussions organized by the UKQCD
\cite{Davies:2002mu}.  Its primary purpose is to facilitate the
advancement of our field by making it easy for scientific
collaborations to share gauge configuration files with other
collaborations throughout the world.  To reach this goal required
first the development of regional or national grids for intraregional
file sharing and storage.  This effort typically exploited existing
grid infrastructure and tools.  The ILDG was then built upon the
regional grids.

As a consequence of this history, the ILDG is a consortium of
autonomous regional grids \cite{Ukawa:2004he}.  Current participants
are centered in Australia (CSSM), Europe (Austria, Switzerland,
Germany, France, Italy)(LDG), Japan (JLDG), the United Kingdom
(QCDGrid or DiGS), and the United States.  Additional participants
are, of course, always welcome.  Cooperation involves the
standardization of several components: binary file formats, metadata
markup schema (QCDml), user authentication methods, and
replica-catalog and file-transfer interoperability.  These standards
are achieved through the efforts of two international working groups,
namely, the Metadata and the Middleware Working Groups.  An ILDG board
with international representation supervises the work of these groups.
The board chair rotates on an annual basis.  To foster coordination,
for the past five years the ILDG has held semiannual internet
conferences through VRVS (Virtual Room Videoconferencing System) with
worldwide participation.

Funding for this effort has been left to countries hosting the
regional grids.  For the most part funding has been minimal if not
completely absent.  The manpower, grid software, and necessary file
storage capacity have been donated and/or borrowed from other funded
projects.  Nonetheless, over the course of five years, the ILDG has
developed into a resource that is ready for use.

This is a brief status report.  Rather than dwelling on the technical
details of grid operation, I will try to take the perspective of a
user.

\section{Access policies}

Lattice files in a grid repository may have access restrictions.
These are set by the contributing collaboration.  Thus it is always a
good idea to check with the collaboration that created the files if
you find there are restrictions.  The authors are often comfortable
releasing recent lattices selectively for noncompeting projects, or
they might be willing to collaborate if there are overlapping
interests.  There is a spectrum of such policies ranging from
permissive to more restrictive:
\begin{itemize}
  \item All lattices are available as soon as they are generated (some
  collaborations).
  \item Recent lattices are available after negotiation (most collaborations).
  \item All lattices are available approximately six months after the
  first publication (universal).
\end{itemize}
Of course, in a publication based on another collaboration's lattices
it is proper to cite an appropriate journal reference, where possible,
and to acknowledge the collaboration that produced them.

It is very encouraging to note a historical trend toward increasing
openness.

\section{Files}

The binary lattice files adhere to a prescribed format \cite{ILDGfmt}
called LIME \cite{LIME}. (The name is a take-off on the name for the
conventional MIME standard for e-mail, in this case meaning ``lattice
interchange message encapsulation'').  The standard gives
collaborations considerable leeway in adding their own variations.
Essentially, the file consists of a series of records (messages), only
three of which have required content: the main one, of course, is the
binary payload.  A second required message specifies the precision and
space-time dimensions.  A third contains the unique logical file name
(LFN), for example,
\begin{verbatim}
lfn://USQCD/MILC/asqtad/2_plus_1_flavor/2064f21b676m007m050/
   series_1/l2064f21b676m007m050b.1530  
\end{verbatim}
The first part of the LFN (\verb|USQCD| in this case) is a
standardized prefix identifying the collaboration, and the rest is
left to the discretion of the collaboration.  

The logical file name (LFN) is used by the catalogs and file transfer
utilities to identify the file.  The mechanism works as follows: a
user requests the file by naming its LFN.  The retrieval software
consults a replica catalog (RC), which associates the LFN with a
universal resource locator (URL).  The URL locates the file on the
internet.

Each binary file is accompanied by a metadata file containing the
archiving history, checksums and plaquette values, etc.  Each ensemble
of gauge configurations also has a metadata file containing
information about the gauge action and the parameters used to generate
it.  Both types of metadata files are XML and must conform to a
standard schema, collectively called QCDml.  The Metadata Working
Group sets all of these standards \cite{QCDML}.  Please see Tomoteru
Yoshie's poster contribution \cite{Yoshie}.

\section{Mode of access}

For a detailed account of file discovery and retrieval, please see
Dirk Pleiter's poster contribution \cite{Pleiter}.

\subsection{Grid certificates}

Before you can download lattices you must obtain an X.509 grid
certificate from an appropriate certificate authority (CA).  The ILDG
accepts certificates from any member of the ILDG Trust Federation
\cite{ILDGCA}. Once you have a certificate, you must register it with
the ILDG ``virtual organization'' \cite{ILDGVOM}.

\subsection{Discovery}

The obvious next step in accessing a set of lattices is to identify
the ensemble(s) of interest.  This is the purpose of the metadata
catalog (MDC).  

Each regional grid maintains a catalog of the files in its repository
and publishes the MDC for those files.  The regional grids then
republish the contents of the world's catalogs through their web
portals.  This process is in various stages of development.
Eventually all the regional grids will catalog the world's ensembles
and be globally searchable.

A web-based global catalog is available in Australia
(CSSM)\cite{CSSM}, Europe (LDG)\cite{LDG}, Japan (JLDG)\cite{JLDG},
and the US (JLab)\cite{JLabMDC}.  Web-based search engines for
searching the MDC's are also available in Australia and Japan.  The UK
provides a Java-based browser application to perform data discovery
\cite{DiGS}.  The United States currently has not populated its own
MDC, but lattices in the older NERSC format continue to be available
by means of a web interface through the Gauge Connection \cite{NERSC},
and Brookhaven National Laboratory \cite{BNL} has recently posted
lattices from the RBC collaboration.

\subsection{Downloading files}

The final step involves the file transfer itself, which can proceed
through various protocols directly (http or gridftp) or through the
Storage Resource Manager (SRM v2) service framework.  The latter
allows client and server to negotiate the transfer protocol and
endpoint that are actually used.

Client software is supposed to support SRM.  The actual transfer
protocol depends on the regional grid that dispenses the files.
Australia, Japan and UK publish their data via a gridftp service. The
JLDG additionally has a convenient web interface that provides the
necessary download script \cite{JLDG}. The LDG dispenses files via SRM
and provides a set of convenient tools called LTOOLS \cite{LDGtools}
to allow integrated access to their metadata, file catalogs, and
storage elements. The UK uses its own set of free grid tools
\cite{DiGS}.  Finally, in the US, NERSC and BNL currently use http,
but produce files only in the old NERSC format.  An initial storage
element is in development at FNAL, where there are now many files in
ILDG format ready for SRM access and cataloging.

The middleware working group is working to alleviate this complex
situation.  Recently packages have been added to the LDG LTOOLS
framework (which is relatively straightforward to install) that
provide an SRM client (srmcp) and an initial version of an integrated
client (ildg-client/ildg-get) that allows retrieval of metadata
documents and files through http or SRM. This software, as well as the
crucial catalogues such as the metadata and file catalogs are in the
process of maturation, and the process should become better integrated
and more reliable in the future.

\section{What ensembles are or will soon be available?}

I give a brief survey of some important and highly useful lattice
ensembles now or soon to be available to the world community.  As
might be expected, they reflect a healthy diversity of approaches to
lattice physics.  I have not attempted an exhaustive list -- rather I
have included what I feel might be particularly interesting. Nor have
I attempted to list all the important lattice parameters, such as
lattice size and quark mass.  The details are available in the meta
data catalogs.  There is such an impressive variety to choose from,
surely there is something for every taste.

\subsection{Australia}

The Adelaide group has only recently gained access to computing
facilities that make it possible to begin serious unquenched
calculations using the fat link irrelevant clover (FLIC) action.
Production has begun as indicated in Table~\ref{listCSSM}.

\begin{table}
\begin{center}
\begin{tabular}{|l|l|l|l|l|l|}
\hline
$N_f$ &
Action &
Collaboration &
a(fm) &
cfgs(approx) &
Availability \\
\hline
0 &
tpLW &
Adelaide &
0.128 &
400 &
Available \\
\hline
2 &
FLIC$^\prime$/tpLW &
Adelaide &
0.125 &
400 $\times$ couple &
In production \\
\hline
\end{tabular}
\end{center}
\caption{Selected ensembles available through the Australian CSSM. The
  abbreviation ``tpLW'' stands for the tadpole L\"uscher-Wei\ss\ 
  action.  For configuration number, the notation $400 \times n$
  implies $n$ ensembles of approximately 400 configurations each,
  typically with different choices of quark masses.
\label{listCSSM}}
\end{table}

\subsection{Europe}

Europe (Table~\ref{listLDG}) has a couple of vigorous dynamical fermion
efforts underway.  The QCDSF collaboration has been generating some
very large two-flavor ensembles based on the nonperturbative clover
action and is following up with $2+1$ flavors of ``stout'' fermions.
The ETMC has also been generating large two-flavor QCD ensembles with
twisted mass fermions and plans to follow up with a full complement of
flavors using a stouted version of the action.

\begin{table}
\begin{center}
\begin{tabular}{|l|l|l|l|l|l|}
\hline
$N_f$ &
Action &
Collaboration &
a(fm) &
cfgs(approx) &
Availability \\
\hline
2 &
NP-clover/Wilson plaq &
QCDSF &
0.07/0.11 &
5000$\times$16 &
Negotiable \\
\hline
2 &
tmQCD/tlSym &
ETMC &
0.07/0.11 &
3500$\times$15 &
Negotiable \\
\hline
$2+1$ &
stout/tpLW &
QCDSF &
0.08 &
1000$\times$16 &
In progress \\
\hline
$2+1+1$ &
Stouted tmQCD/tlSym &
ETMC &
0.07/0.09 &
2500$\times$ 15 &
In progress \\
\hline
\end{tabular}
\end{center}
\caption{Selected ensembles available through the European LDG.
  Notation is the same as in previous tables. The abbreviation
  ``tlSym'' stands for the tree-level Symanzik gauge action.  The
  abbreviation ``tmQCD'' stands for the twisted mass formulation.
\label{listLDG}
}
\end{table}

\subsection{Japan}

In Japan (Table~\ref{listJLDG}) the CP-PACS and JLQCD collaborations
have access to very substantial computational resources, which have
been devoted to two ambitious lattice generation projects using clover
fermions and overlap fermions.  The CP-PACS and JLQCD collaborations
have recently released a significant set of full-QCD ensembles based
on the Wilson-clover fermion action with the Iwasaki gauge action.

\begin{table}
\begin{center}
\begin{tabular}{|l|l|l|l|l|l|}
\hline
$N_f$ &
Action &
Collaboration &
a(fm) &
cfgs(approx) &
Availability \\
\hline
&
&
&
0.20 &
1000$\times$4 &
\\
2 &
Clover/Iwasaki &
CP-PACS &
0.15 &
1000$\times$4 &
Available \\
&
&
&
0.10 &
800$\times$4 &
\\
\hline
2 &
Clover/Plaquette &
JLQCD &
0.10 &
1200$\times$5 &
In preparation
\\
\hline
&
&
&
0.12 &
800$\times$5$\times$2 &
\\
$2+1$ &
Clover/Iwasaki &
CP-PACS+&
0.10 &
800$\times$5$\times$2 &
Available (new)
\\
&
&
JLQCD&
0.07 &
600$\times$5$\times$2 &
\\
\hline
2 &
Overlap/Iwasaki &
JLQCD &
0.12 &
O(500)$\times$6 &
After 1st spectrum paper
\\
\hline
\end{tabular}
\end{center}
\caption{Selected ensembles available through the Japanese JLDG. 
  Notation is the same as in previous tables.
\label{listJLDG}
}
\end{table}

\subsection{UKQCD and RBC Collaborations}

Using QCDOC computers, the UKQCD collaboration has teamed up with the
US RBC collaboration to generate a series of lattice ensembles based
on the domain wall fermion formulation (Table~\ref{listUKQCD}).  A
large set has been released for general public access.  A few of the
most recent ensembles still have restricted access.  The UKQCD has
also been generating some full QCD lattices using the Asqtad
formulation.

\begin{table}
\begin{center}
\begin{tabular}{|l|l|l|l|l|l|}
\hline
$N_f$ &
Action &
Collaboration &
a(fm) &
cfgs(approx) &
Availability \\
\hline
$2+1$ &
DWF/Iwasaki &
UKQCD/RBC &
0.12 &
800x4 &
Available \\
\hline
$2+1$ &
DWF/DBW2 &
UKQCD/RBC &
0.12 &
300-1000 &
Available \\
\hline
2 &
DWF/DBW2 &
UKQCD/RBC &
0.12 &
1000x3 &
Available \\
\hline
$2+1$ &
Asqtad/tpLW &
UKQCD &
0.12 &
3000 &
Available \\
\hline
$2+1$ &
DWF/Iwasaki &
UKQCD/RBC &
0.09 &
?? &
Restricted \\
\hline
$2+1$ &
DWF/Iwasaki &
UKQCD/RBC &
0.09 &
?? $\times$ 2 &
In production \\
\hline
\end{tabular}
\end{center}
\caption{Selected ensembles available through the UKQCD and RBC
  collaborations.  Notation is the same as in previous tables.  The
  question mark indicates the information was unknown at the time of
  this writing.
\label{listUKQCD}
}
\end{table}

\subsection{United States}

In the US (Table~\ref{listUS}) over the past few years the MILC
collaboration has been generating and publishing an extensive set of
full QCD ensembles based on the Asqtad formulation.  The most recent
additions are at a lattice spacing of 0.06 fm.  Further work is
underway to extend and add to the existing ensembles at 0.09 and 0.12
fm.  All lattices are open to general access as soon as they are
created.  The LHPC collaboration has embarked on a major effort to
generate dynamical clover ensembles on anisotropic lattices.  The US
RBC production of domain wall lattices was included in the previous
section.

\begin{table}
\begin{center}
\begin{tabular}{|l|l|l|l|l|l|}
\hline
$N_f$ &
Action &
Collaboration &
a(fm) &
cfgs(approx) &
Availability \\
\hline
$2+1$ &
 Asqtad/tpLW &
MILC &
0.15 &
600 x 4 &
Available \\
\hline
$2+1$ &
Asqtad/tpLW &
MILC &
0.12 &
600 x 5 &
Available \\
\hline
$2+1$ &
Asqtad/tpLW &
MILC &
0.09 &
500$\times$2 &
Available \\
\hline
&
&
&
&
500 $\times$ 1 &
New, available \\
\hline
$2+1$ &
Asqtad/tpLW &
MILC &
0.06 &
500 $\times$ 2&
In production \\
&
&
&
&
&
Available \\
\hline
$2+1$ &
Aniso clover/tpLW &
LHPC &
0.125 &
500 $\times$ 4 &
Planned \\
\hline
$2+1$ &
Aniso clover/tpLW &
LHPC &
0.10 &
?? $\times$ 4 &
Planned \\
\hline
$2+1$ &
Aniso clover/tpLW &
LHPC &
0.08 &
?? $\times$ 4 &
Planned \\
\hline
\end{tabular}
\end{center}
\caption{Selected ensembles available through the NERSC gauge
  connection or by contacting the collaborations directly.
  Notation is the same as in previous tables.
\label{listUS}
}
\end{table}

\acknowledgments
I am indebted to Balint Joo and Dirk Pleiter for checking over and
contributing portions of the description of the middleware software
operation.

\end{document}